# Acoustic Geometric-Phase Meta-Array

Bingyi Liu[1], Zhaoxian Su[1], Yong Zeng[2], Yongtian Wang[1], Lingling Huang[1,*] and Shuang Zhang[3,4†]

[1]*School of Optics and Photonics, Beijing Engineering Research Center of Mixed Reality and Advanced Display, Beijing Institute of Technology, Beijing, 100081, China*
[2]*Faculty of Materials and Manufacturing, Beijing University of Technology, Beijing, 100124, China*
[3]*Department of Physics, The University of Hong Kong, Hong Kong, China*
[4]*Department of Electrical and Electronic Engineering, The University of Hong Kong, Hong Kong, China*

## Abstract

Metasurfaces based on geometric phase acquired from the conversion of the optical spin states provide a robust control over the wavefront of light, and have been widely employed for construction of various types of functional metasurface devices. However, this powerful approach cannot be readily transferred to the manipulation of acoustic waves because acoustic waves do not possess the spin degree of freedom. Here, we propose the concept of acoustic geometric-phase meta-array by leveraging the conversion of orbital angular momentum of acoustic waves, where well-defined geometric-phases can be attained through versatile topological charge conversion processes. This work extends the concept of geometric-phase metasurface from optics to acoustics, and provides a new route for acoustic wave control.

---

[*] Email: huanglingling@bit.edu.cn
[†] Email: shuzhang@hku.hk

Geometric-phase metasurfaces (GPMs) have been widely employed as an efficient and robust means to control the scattering of electromagnetic waves with the controllable geometric phase acquired via a spin conversion process [1-3]. Due to the simple relationship between the geometric-phase and the orientation of anisotropic nanostructures for generating the spin conversion, GPMs can be used for flexible wavefront manipulation, such as imaging [4,5], holography [6,7], harmonic generations [8,9], trapping [10,11], quantum technology [12,13], etc. However, this concept cannot be directly extended to acoustics due to the lack of such spin conversion process in acoustics. Therefore, the common strategies involved in the design of acoustic metasurfaces or metamaterials mainly focus on acoustic surface impedance engineering and effective acoustic refractive index modulation, such as Helmholtz resonators [14-17], grooves [18-20], coiling-space structures [21-25], pentamode metamaterials [26], mass-membrane system [27], just to name a few. In addition, most of the reported acoustic meta-devices can only support a given acoustic wave modulation, i.e., the functionality of acoustic meta-device is fixed once the geometry of the structures is determined. Hence, acoustic metasurfaces usually require precise spatial arrangement of the meta-atoms by following a rigorous meta-device design procedure [28]. As a result, recent works on tunable acoustic meta-atoms suffer from the resonance-based phase delay, which unavoidably couples with the transmission amplitude and requires delicate control over some critical geometry parameters [29-31].

In this work, we propose an acoustic meta-array based on acoustic geometric phases which are obtained through versatile acoustic vortex topological charge (TC) conversion occurring within each pixel of the meta-array—a cylindrical acoustic waveguide with judiciously engineered interior structures. Each meta-pixel waveguide, containing a number of acoustic geometric meta-plates (AGM), is designed to implement sequential manipulations on the incident acoustic wave including generation and conversion of vortex beams of various TCs. A geometric phase arises from the conversion between two different orbital angular momenta, which can be continuously controlled by varying the orientation of the conversion element, i.e., the AGM for the designated TC conversion process. As a proof of principle, flexible acoustic field manipulations including acoustic beam steering and focusing can be realized with the proposed geometric-phase meta-array. Our design opens a new avenue for flexible acoustic field generation and the controllable phase manipulation.

We start by a brief description of the geometric phase with optical metasurfaces. When a circularly polarized light is incident onto a geometric-phase metasurface, the transmitted light carries a geometric phase term of $\exp[(\sigma^{in} - \sigma^{out})\theta_\sigma]$, where $\sigma^{in}$ and $\sigma^{out}$ refer to the spin of input and output circularly polarized light, $\theta_\sigma$ is the orientation angle of the nanoantenna, as illustrated by Fig. 1(a). Therefore, only the cross-polarized component of the transmitted light, i.e., $\sigma^{out} = -\sigma^{in}$, carries the geometric phase term of $\exp(2i\sigma^{in}\theta_\sigma)$. Thus, the wavefront of the cross-polarized transmitted beam can be arbitrarily controlled by adjusting the orientation angles of the nanoantennas across the metasurface. Here, we show that this geometric phase also accompanies a conversion process between different orbital angular momenta, and therefore the same principle can be applied to acoustics where spin degree of freedom does not exist. Specifically, when an acoustic meta-plate converts the vortex of an incident acoustic beam from TC $l^{in}$ to TC $l^{out}$, the TC-converted acoustic beam could carry an additional phase modulation of $\exp[i(l^{in} - l^{out})\theta_l]$, where $\theta_l$ is the orientation angle of the acoustic meta-plate responsible for this conversion, as illustrated by Fig. 1(b).

In order to understand the mechanism of the presence of the acoustic geometric phase obtained through TC conversions, we consider a general TC conversion case which involves the orbital angular momentum transfer from TC $l^{in}$ to TC $l^{out}$. By defining the complex transmission of an AGM of TC $l^\xi$ and orientation angle $\theta_l$ as $\hat{T}(\theta_l)$, then we have:

$$|l^{out}\rangle = \hat{T}(\theta_l)|l^{in}\rangle \quad (1)$$

By rotating the incident beam, the transmission beam, and the AGM in the counter-clockwise direction about the propagation axis by an angle $\varphi$, the above equation still holds for the transformed waves. Considering the additional phase acquired by the incident and transmitted beams due to the rotation (see Section I of Supplementary Material [32]), one can write:

$$\exp(-il^{out}\varphi)|l^{out}\rangle = \hat{T}(\theta_l + \varphi)\exp(-il^{in}\varphi)|l^{in}\rangle \quad (2)$$

By combing Eqn. 1 and Eqn.2, it is straightforward to show $\hat{T}(\theta_l + \varphi) = \hat{T}(\theta_l)\exp[i(l^{in} - l^{out})\varphi]$, and by selecting $\theta_l$ to be zero, we arrive at the following simple expression for the geometric phase as a function of orientation angle $\varphi$:

$$\hat{T}(\varphi) = \hat{T}(0)\exp[i(l^{in} - l^{out})\varphi] \quad (3)$$

Hence, the overall geometric phase carried by the transmitted field is $\exp[i(l^{in} - l^{out})\varphi]$.

Thus, various types of geometric-phase modulations are available via different TC conversion processes. For example, for an incident beam of TC $l^{in} = 1$, with an acoustic meta-plate of TC $l^\xi = 1$, the transmitted acoustic wave would have a TC $l^{out} = 0$ and carry the geometric phase of $\exp(i\theta_l)$ when rotating the acoustic meta-plate by angle of $\theta_l$. Moreover, by converting an acoustic vortex beam of TC $l^{in} = \pm 1$ to the vortex of TC $l^{out} = \mp 1$ through an acoustic meta-plate of TC $l^\xi = 2$, the corresponding geometric-phase modulation of transmitted acoustic wave is $\exp(\pm 2i\theta_l)$. Generally, the acoustic geometric phase $\exp(qi\theta_l)$, $q = \pm 1, \pm 2 \cdots$ can be obtained at will by selecting appropriate TC conversion process.

Ideally, an AGM can be constructed by dividing it into many sections along the azimuthal direction and by filling different sections with acoustic material of gradually varying refractive indices but with the same impedance matched to that of air, as shown in Fig. 2(a). Here, the AGM is placed inside a rigid waveguide of inner radius $R$, and it contains $l^\xi$ repeating units along the azimuthal direction. Each unit consists of $M$ number of sections to achieve a phase variation of $2\pi$ at the operating wavelength $\lambda_0$. Thus, the refractive index of the $m$-th section within each repeating unit is given as $n_m = 1 + (m-1)\lambda_0/Mh$, where $h$ is the thickness of AGM and $m = 1, 2, \cdots, M$, with the corresponding sector angle of each section being $2\pi/(l^\xi M)$.

The numerical calculation is conducted with the finite element method (FEM). Fig. 2(b c) show the geometric-phase modulation obtained through the TC conversion processes. Here, the acoustic meta-plate consists of 12 sections, and the operating wavelength is selected as 10 cm. In the first case shown in Fig. 2(b), the vortex of TC $l = 1$ is firstly generated through $|0\rangle \rightarrow |1\rangle$ process via an AGM of TC $l^\xi = -1$, and then converted into a plane acoustic wave via $|1\rangle \rightarrow |0\rangle$ process by a second AGM of TC $l^\xi = +1$. The orientation of the first AGM is fixed while that of the second one ($\theta_l$) is varied to provide a geometric phase of $\exp(i\theta_l)$ (Fig. 2(d)). In the second example, the geometric phase obtained through the TC conversion process $|1\rangle \to |-1\rangle$ is illustrated in Fig. 2(c). In this configuration, three AGMs are employed, i.e., vortex source generation (AGM1, $l^\xi = 1$, $|0\rangle \to |1\rangle$), TC conversion (AGM2, $l^\xi = 2$, $|1\rangle \to |-1\rangle$) and conversion back into a plane wave (AGM3, $l^\xi = 1$, $|-1\rangle \to |0\rangle$), see Fig. 2(c). The geometric phase retrieved from the simulation is proportional to $\exp(2i\theta_l)$ where $\theta_l$ is the orientation of the 2$^{nd}$ AGM, as shown in Fig. 2(e). The geometric-phase modulation obtained through higher order TC conversions, such as $|2\rangle \to |-1\rangle$ ($\propto \exp(3i\theta_l)$) and $|3\rangle \to |-1\rangle$ ($\propto \exp(4i\theta_l)$), are shown in Section II and III of Supplementary Material [32].

To showcase the potential of acoustic geometric phase in wavefront manipulation, we theoretically study the beam steering and beam focusing by using the ideal geometric-phase meta-arrays described above. Fig. 3(a) shows the full-wave simulation of the beam steering via $|1\rangle \to |0\rangle$. The geometric-phase meta-array consists of 12 pixels of TC $l^\xi = 1$. The operating wavelength $\lambda$ is selected as 12 cm, the pixel size $p_{meta}$ is 9 cm, which is slightly larger than the outer diameter of waveguide (8 cm). The orientation angle $\theta_l$ of AGM2 is varied linearly along $x$ direction, with a uniform step of 30°. The anomalous refracted angle retrieved from the calculated far-field is $-6.3°\pm0.1°$, which agrees well with the theoretical value obtained based on the geometric phase of $\theta_l$. For the $|1\rangle \to |-1\rangle$ TC conversion process, the anomalous refracted angle obtained from the simulation is $-12.8°\pm0.1°$ [Fig. 3(b)], which agrees well with the theoretical value calculated according to the geometric phase of $2\theta_l$. Interestingly, by inverting the TC of AGM1 and AGM3 from $l^\xi = 1$ to $l^\xi = -1$, the geometric phase inverts its sign. We further show beam focusing realized by an acoustic geometric-phase metalens via $|1\rangle \to |0\rangle$ process, as shown by the left panel of Fig. 3(c). The phase profile for focusing in the $x$-$z$ plane is $\varphi(x) = -k(f_z - \sqrt{x^2 + f_z^2})$, where $k = 2\pi/\lambda$ is the operating wavenumber, the designed focal length is $10\lambda$ and the orientation angle of AGM2 in each pixel follows exactly $\varphi(x)$. The top- and bottom-right panel of Fig. 3(c) show the field intensity profile along two perpendicular lines A and B across the focal point. The simulated focal length is about 1.16 m which is close to the designed value 1.2 m. Besides the focusing of free-space plane acoustic waves, other beam manipulations can also be readily achieved.

Next, we utilize realistic acoustic meta-structures to achieve wave manipulation based on the acoustic geometric phase. Here, the acoustic meta-atoms designed for different sections are classical Helmholtz resonator-straight pipe hybrid structures [33], as shown in Fig. 4(a). By tuning the widths of the open pipe ($u_a$, $u_b$, $u_c$) and the open neck ($v_a$, $v_b$, $v_c$) of the Helmholtz resonators of each layer, a high transmission can be achieved. Fig. 4(a) shows the design of two meta-plates of TC $l^\xi = 1$ and $l^\xi = 2$, with the detailed geometry parameters of the constituent units given in Section IV of Supplementary Material [32]. Here, AGMs of TC $l^\xi = 1$ and $l^\xi = 2$ are numerically investigated, and the optimized working frequency in our design is 2.88 kHz. Fig. 4(b) shows the TC generation, conversion and detection with the above two types of AGMs, where the acoustic vortex sources are obtained by illuminating plane acoustic waves onto the meta-plate of TC $l^\xi = 1$. Fig. 4(c) and (d) show the acoustic geometric phase (red solid triangle) and the corresponding transmitted amplitude (blue solid circle) for two systems with cascaded AGMs for realizing geometric phase based on $|1\rangle \rightarrow |0\rangle$ and $|1\rangle \rightarrow |-1\rangle$ TC conversion processes, respectively. It is obvious that the geometric-phase modulation obtained with the realistic acoustic meta-plates agrees well with theoretical predictions. However, due to the multiple scattering induced by the realistic AGMs, the transmitted field uniformity and transmission efficiency would be deteriorated when more AGMs are cascaded. In our simulation, the distance between two meta-plates in $|1\rangle \rightarrow |0\rangle$ process shown in Fig. 4(c) and the interval among three meta-plates in $|1\rangle \rightarrow |-1\rangle$ process shown in Fig. 4 (d) are optimized as 30 cm and 10 cm, respectively. It should be noted that the influence of thermal viscosity on the acoustic geometric-phase modulation is negligible (see Section V of Supplementary Material [32]). In Section VI of Supplementary Material [32], we provide the full-wave simulation of broadband beam bending (from 1.9 kHz to 3.1 kHz) realized by a geometric-phase meta-array made up of the realistic acoustic meta-plates. It is found that our geometric-phase acoustic meta-array could operate in a broadband frequency range, benefitting from the dispersionless character of geometric phase.

In summary, we have proposed the concept of acoustic geometric phase generated in the process of TC conversions, which is a generalization of optical geometric phase for spin conversion. Different from the conventional resonance-type phase modulation, which relies on the geometry and usually tangled with transmission amplitude, acoustic geometric phase is highly robust, and it enables arbitrary phase manipulation by simply rotating the meta-plates. Our work transfers the concept of optical geometric phase to acoustics, and shows its potential in constructing broadband and reconfigurable acoustic meta-devices for arbitrary field generations and manipulations.

## References

[1] E. Hasman, V. Kleiner, G. Biener and A. Niv, Appl. Phys. Lett. **82**, 328 (2003).

[2] D. Lin, P. Fan, E. Hasman and M. L. Brongersma, Science **345**, 298 (2014).

[3] S. Xiao, J. Wang, F. Liu, S. Zhang, X. Yin and J. Li, Nanophotonics **6**, 1 (2017).

[4] M. Khorasaninejad, W. T. Chen, R. C. Devlin, J. Oh, A. Y. Zhu and F. Capasso, Science **352**, 6290 (2016).

[5] S. Wang, P. C. Wu, V.-C. Su, Y.-C. Lai, M.-K Chen, H. Y. Kuo, B. H. Chen, Y. H. Chen, T.-T. Huang, J.-H. Wang, R.-M. Lin, C.-H. Kuan, T. Li, Z. Wang, S. Zhu and D. P. Tsai, Nat. Nanotechnol. **13**, 3 (2018).

[6] L. Huang, X. Chen, H. Mühlenbernd, H. Zhang, S. Chen, B. Bai, Q. Tan, G. Jin, K.-W. Cheah, C.-W. Qiu, J. Li, T. Zentgraf and S. Zhang, Nat. Commun. **4**, 2808 (2013).

[7] G. Zheng, H. Mühlenbernd, M. Kenney, G. Li, T. Zentgraf, and S. Zhang, Nat. Nanotechnol. **10**, 4 (2015).

[8] S. Chen, G. Li, F. Zeuner, W. H. Wong, E. Y. B. Pun, T. Zentgraf, K. W. Cheah and S. Zhang, Phys. Rev. Lett. **113**, 033901 (2014).

[9] M. Tymchenko, J. S. Gomez-Diaz, J. Lee, N. Nookala, M. A. Belkin and A. Alù, Phys. Rev. Lett. **115**, 207403 (2015).

[10] T. Chantakit, C. Schlickriede, B. Sain, F. Meyer, T. Weiss, N. Chattham and T. Zentgraf, Photonics Res. **8**, 1435 (2020).

[11] L. Zhu, X. Liu, B. Sain, M. Wang, C. Schlickriede, Y. Tang, J. Deng, K. Li, J. Yang, M. Holynski, S. Zhang, T. Zentgraf, K. Bongs, Y.-H. Lien and G. Li, Sci. Adv. **6**, eabb6667 (2020).

[12] K. Wang, J. G. Titchener, S. S. Kruk, L. Xu, H.-P. Chung, M. Parry, I. I. Kravchenko, Y.-H. Chen, A. S. Solntsev, Y. S. Kivshar, D. N. Neshev, A. A. Sukhorukov, Science **361**, 1104 (2018).

[13] P. Georgi, M. Massaro, K.-H Luo, B. Sain, N. Montaut, H. Hermann, T. Weiss, G. Li, C. Silberhorn and T. Zentgraf, Light Sci. Appl. **8**, 70 (2019).

[14] S. Zhang, L. Yin and N. Fang, Phys. Rev. Lett. **102**, 194301 (2009).

[15] Y. Li, S. Qi and M B. Assouar, New J. Phys, **18**, 043024 (2016).

[16] Y. Fu, C Shen, X. Zhu, Y. Liu, Y. Liu, S. A. Cummer, Sci. Adv. **6**, eaba9876 (2020).

[17] Y. Zhu, X. Fan, B. Liang, J. Cheng and Y. Jiang, Phys. Rev. X **7**, 021034 (2017).

[18] J. Zhao, B. Li, Z. Chen amd C.-W. Qiu, Appl. Phys. Lett. **103**, 151604 (2013).

[19] X. Wu, X. Xia, J. Tian, Z. Liu and W. Wen, Appl. Phys. Lett. **108**, 163502 (2016).

[20] A. Díaz-Rubio, J. Li, C. Shen, S. A. Cummer and S. A. Tretyakov, Sci. Adv. **5**, eaau7288 (2019).

[21] Z. Liang and J. Li, Phys. Rev. Lett. **108**, 114301 (2012).

[22] Y. Li, B. Liang, X. Tao, X. Zhu, X. Zou and J. Cheng, Appl. Phys. Lett. **101**, 233508 (2012).

[23] Y. Xie, W. Wang, H. Chen, A. Konneker, B.-I Popa and S. A. Cummer, Nat. Commun. **5**, 5553 (2014).

[24] X. Zhu, K. Li, P. Zhang, J. Zhu, J. Zhang, C. Tian and S. Liu, Nat. Commun. **7**, 11731 (2016).

[25] H. Esfahlani, H. Lissek and J. R. Mosig, Phys. Rev. B, **95**, 024312 (2017).

[26] Y. Tian, Q. Wei, Y. Cheng, Z. Xu, and X. Liu, Appl. Phys. Lett. **107**, 221906 (2015).

[27] G. Ma, M. Yang, S. Xiao, Z. Yang and P. Sheng, Nat. Mater. **13**, 9 (2014).

[28] G. Memoli, M. Caleap, M. Asakawa, D. R. Sahoo, B. W. Drinkwater and S. Subramanian, Nat. Commun. **8**, 14608 (2017).

[29] S. Zhao, A. Chen, Y. Wang and C. Zhang, Phys. Rev. Appl. **10**, 054066 (2018).

[30] Z. Tian, C. Shen, J. Li, E. Reit, Y. Gu, H. Fu, S. A. Cummer, T. J. Huang, Adv. Funct. Mater. **29**, 13 (2019).

[31] C. Zhang, W. K. Cao, L. T. Wu, J. C. Ke, Y. Jing, T. J. Cui and Q. Cheng, Appl. Phys. Lett. **118**, 133502 (2021).

[32] See Supplementary Material for additional discussions on the geometric-phase obtained by rotating the acoustic vortex sources, acoustic geometric-phase modulation obtained with phased microphone illumination, geometry parameters of meta-plate and the influence of thermal viscosity on the performance of designed geometric-phase meta-plate, broadband acoustic beam bending realized with real geometric-phase meta-array.

[33] X. Jiang, Y. Li, B. Liang, J. -C. Cheng and L. Zhang, Phys. Rev. Lett. **117**, 034301 (2016).

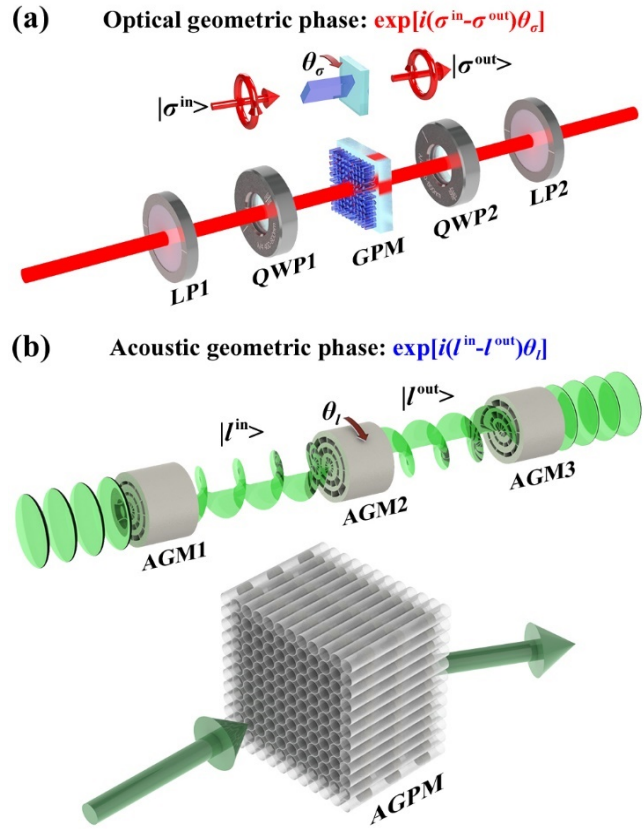

FIG. 1. Analogy between the (a)optical and (b)acoustic geometric phase obtained through scattering processes in meta-atoms. LP, QWP and GPM refer to the linear polarizer, quarter wave plate and optical geometric-phase metasurface, respectively. AGMs refer to the acoustic meta-plates designed for TC generation, conversion and detection. The acoustic geometric-phase meta-array (AGPM) consists of cylindrical waveguides to support the TC conversion processes.

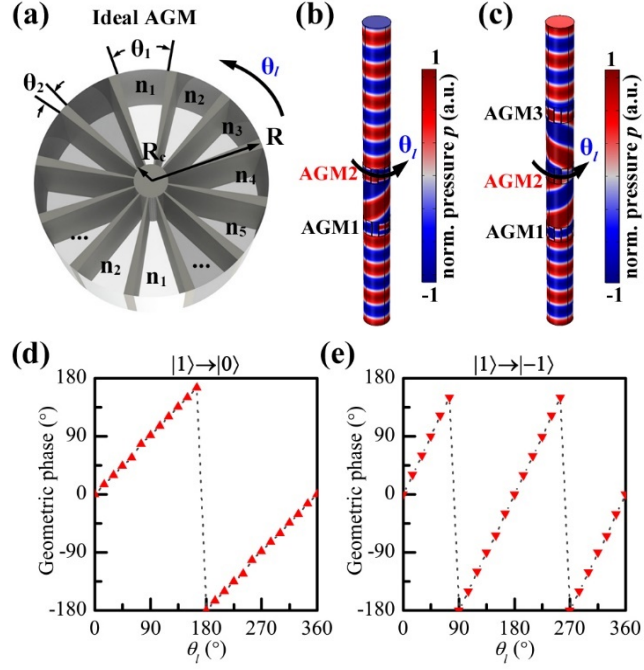

FIG. 2. Geometric-phase modulation defined by TC conversion in acoustic meta-plates. Schematic of (a)ideal acoustic meta-plate, the radius of waveguide $R$ and solid center $R_c$ is $0.4\lambda$ and $0.05\lambda$, the thickness $h$ of AGM is $0.5\lambda$, the section angle $\theta_1$ and $\theta_2$ is 27° and 3°, respectively. Configuration of the cascaded acoustic meta-plates for geometric phase encoding process (b)$|0\rangle \to |1\rangle \to |0\rangle$ and (c)$|0\rangle \to |1\rangle \to |-1\rangle \to |0\rangle$, and the corresponding geometric phase obtained via TC conversion processes vs the orientation angle of AGM2 for (d)$|1\rangle \to |0\rangle$ and (d)$|1\rangle \to |-1\rangle$.

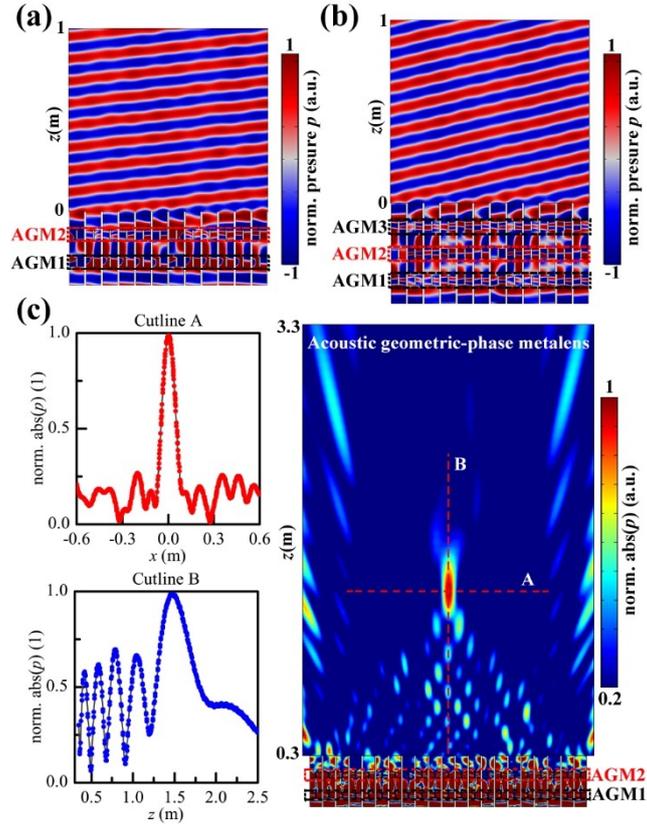

FIG 3. Acoustic wave manipulation based on geometric phase. Beam steering via (a)$|1\rangle \to |0\rangle$ and (b)$|1\rangle \to |-1\rangle$ TC conversion process. (c)Focusing of free-space plane acoustic waves with geometric-phase meta-array by utilizing $|1\rangle \to |0\rangle$ TC conversion process.

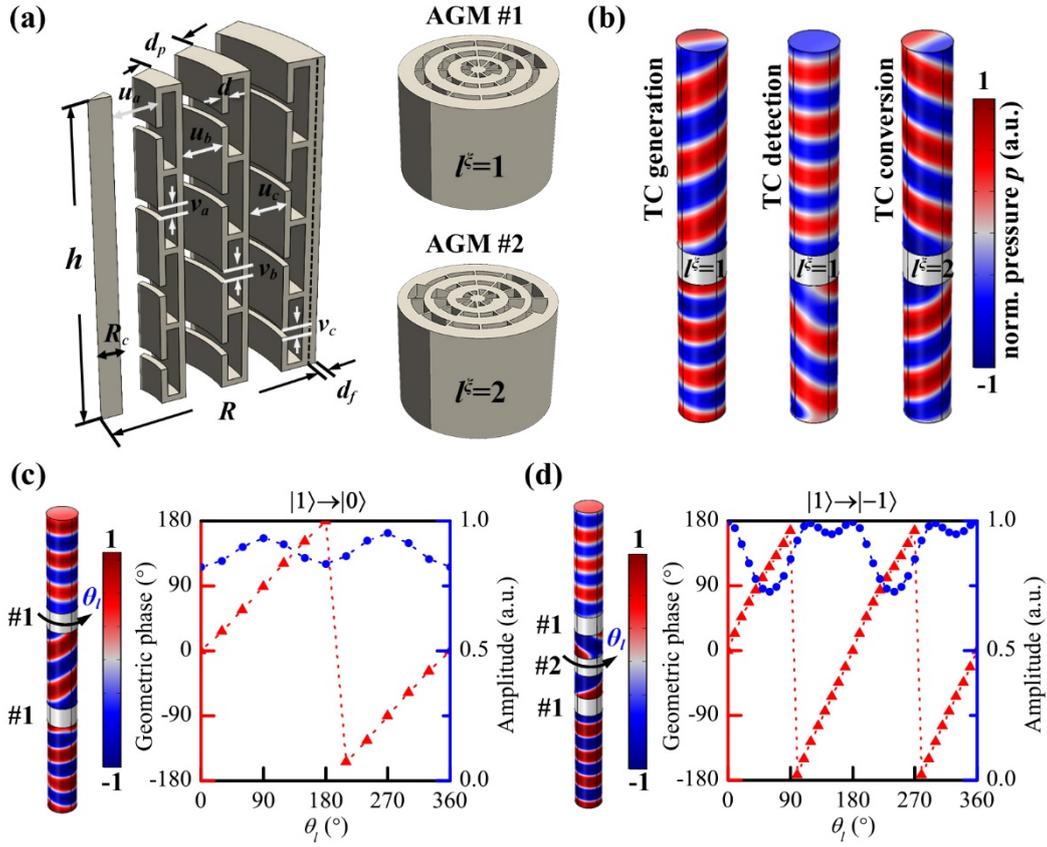

FIG. 4. Realistic acoustic meta-structures for implementing acoustic geometric phases. (a)Schematic of the designed meta-plate. The height $h$ and radius $R$ of designed meta-plate are 7.5 cm and 5 cm, respectively. The thickness of the rigid wall $d$ is 1.5 mm, the period of each structure layer along the radial direction $d_p$ is 1.5 cm, the radius $R_c$ of rigid center is 3 mm. An additional outer rigid wall $d_f$ of thickness 2 mm is added to fit the waveguide. (b)Full-wave simulation of realistic meta-plates for TC generation (left), detection (middle) and conversion (right). Geometric phase (red solid triangle) and transmission spectrum (blue solid circle) obtained with TC conversion processes (c)$|1\rangle \to |0\rangle$ and (d)$|1\rangle \to |-1\rangle$.